\documentclass[aoas,nameyear,seceqn,dvips]{arximspdf}
\usepackage{graphics}
\doi{10.1214/09-AOAS324}
\volume{4}
\issue{3}
\pubyear{2010}
\firstpage{1430}
\lastpage{1450}

\makeatletter
\def\eqref#1{(\ref{#1})}

\newproclaim{example}{Example}
\makeatother

\begin{document}
\begin{frontmatter}

\title{Optimal designs for random effect models with correlated errors
with applications in~population~pharmacokinetics\thanksref{T1}}
\runtitle{Optimal designs for random effect models}
\thankstext{T1}{Supported by Deutsche Forschungsgemeinschaft (SFB 823,
``Statistik nichtlinearer dynamischer Prozesse.'' Teilprojekt C2).}

\begin{aug}
\author[A]{\fnms{Holger} \snm{Dette}\thanksref{t2}\ead[label=e1]{holger.dette@rub.de}\corref{}},
\author[B]{\fnms{Andrey} \snm{Pepelyshev}\ead[label=e2]{andrey@ap7236.spb.edu}} \and
\author[C]{\fnms{Tim} \snm{Holland-Letz}\ead[label=e3]{tim.holland-letz@rub.de}}
\thankstext{t2}{Supported in part by NIH Grant IR01GM072876:01A1 and the BMBF project SKAVOE.}
\runauthor{H. Dette, A. Pepelyshev and T. Holland-Letz}
\affiliation{Ruhr University at Bochum, St. Petersburg State University
and Ruhr~University~at~Bochum}
\address[A]{H. Dette\\ Fakult\"at f\"ur Mathematik \\ Ruhr-Universit\"
at Bochum \\
44780 Bochum\\ Germany \\
\printead{e1}} 
\address[B]{A. Pepelyshev \\
Department of Mathematics \\
St. Petersburg State University \\
St. Petersburg\\ Russia \\
\printead{e2}}
\address[C]{T. Holland-Letz \\
Medizinische Fakult\"at \\
Ruhr-Universit\"at Bochum \\
Abteilung f. medizinische Informatik \\
44780 Bochum\\ Germany \\
\printead{e3}}
\end{aug}

\received{\smonth{8} \syear{2009}}
\revised{\smonth{12} \syear{2009}}

%
\begin{abstract}
We consider the problem of constructing optimal designs for population
pharmacokinetics which use random effect models. It is common
practice in the design of experiments in such studies to assume
uncorrelated errors for each subject. In the present paper a
new approach is introduced to determine efficient designs for nonlinear
least squares estimation which addresses the problem of correlation between
observations corresponding to the same subject. We
use asymptotic arguments to derive optimal design densities, and the
designs for finite sample sizes are constructed from the quantiles of the
corresponding optimal distribution function. It is demonstrated that
compared to the optimal exact designs, whose determination is a hard
numerical problem,
these designs are very efficient.
Alternatively, the designs derived from asymptotic theory could be used
as starting designs for the numerical computation of exact optimal
designs. Several examples of linear and nonlinear models are presented
in order
to illustrate the {methodology}. In particular, it is demonstrated that
naively chosen equally spaced
designs may lead to less accurate estimation.

\end{abstract}

%

\end{frontmatter}

\section{Introduction}\label{sec1}

The~work presented in this paper
is motivated by some problems encountered in the design of experiments
in a clinical trial
to establish the pharmacokinetics of Uzara\tsup{\textregistered}, a digitoxin
related herbal diarrhea medication [based on \citet{thuermann2004}].
These kinds of trials pose methodological design challenges because
they require the estimation of global population parameters in the
presence of
correlated measurement errors.
The~trial in question included a number of patients each given an oral
application of Uzara\tsup{\textregistered} as well as, after a washout
period, an intravenous application of digoxin (Lanicor\tsup{\textregistered}), where in both cases the resulting serum concentration of digitoxin
was measured repeatedly during the next $36$ hours.

The~relation
between the time and the concentration in the analysis of the
Uzara\tsup{\textregistered}\ trial
can be described using the theory of one-compartment models with oral
and, respectively, intravenous applications [\citet{atchheju1993},
\citet{shargel1993}]. In the intravenous case, the medication reaches
the maximum concentration in the blood almost immediately, and, after
that, it is gradually eliminated from the body over time. Thus, the
digitoxin concentration is modeled using the exponential elimination model
%
\begin{eqnarray}
\label{2par} \eta(t,b)=b_1 e^{-b_2 t}.
\end{eqnarray}
In the case of the oral application, there is a gradual build up of the
concentration in the blood as the medication is absorbed through the
digestive tract, while there is a simultaneous elimination process of
the medication in the blood. Therefore, the concentration function is
the solution of the differential equation of these two parallel
processes. The~resulting function
%
\begin{eqnarray}
\label{3pramBat} \eta(t,b)=b_3(e^{-b_1 t}-e^{-b_2 t})
\end{eqnarray}
is known as the ($3$ parameter) Bateman function [\citet{garrett1994}].
In both models, the function $\eta(t,b)$ denotes the concentration function,
$t$ is the time (in hours), $b=(b_1,b_2)$ and, respectively, $b=(b_1,b_2,b_3)$
are the vectors of parameters. The~parameters are assumed to vary
between patients and
the aim of the experiment is to estimate their global
means (and sometimes variances) over all patients.

Measurements within the same patient are usually correlated, and we
assume this correlation to be
proportional to the time lag between measurements, which is plausible
as the random errors are usually caused by temporary changes in the
patients physical condition.
Measurements for
different patients are assumed to be independent.
In the trial at question Thuermann considered $n=15$ (oral),
respectively, $n=14$ (intravenous) measurements each on $K=18$ patients.
After a preliminary discussion
with experts the measurements
were taken at nonoptimized time points.
An approximation
of the covariance of a single patient can then be expressed as
%
\begin{eqnarray}\label{covpop}
\Sigma_{\mathit{pop}}=\frac{\partial\eta(\mathbf{t},b)}{\partial b}^T
V_p\frac{\partial\eta(\mathbf{t},b)}{\partial b}+
V_\varepsilon,
\end{eqnarray}
where $\eta(\mathbf{t},b)=(\eta(t_1,b),\ldots,\eta(t_n,b))^T$ denotes
the vector of expected responses at $t_1,\ldots,t_n$ and $V_\varepsilon$
is the covariance matrix corresponding to this data. This expression
includes two sources of variation, the usual variation $V_\varepsilon$
caused by random errors as well as the additional variation
$V_p$ due to the random effect.

Situations of this kind are rather common
in the evaluation of the pharmacokinetics and the pharmacodynamics of
drugs [see \citet{buelga2005}, \citet{colombo2006}, among others].
The~corresponding processes are usually modeled by linear or nonlinear
random effects models, which try to estimate the population parameters,
that is, the mean and the inter-individual variability of the
parameters. Under the additional assumption of a normal distribution,
the population characteristics are usually estimated by maximum
likelihood methods. In many cases,
the likelihood cannot be evaluated explicitly and approximations are
used to calculate the estimate. Efficient
algorithms for estimation are available for this purpose [see \citet
{aarons1999}].
Loosely speaking, under a Gaussian assumption this approach corresponds to
weighted nonlinear least squares estimation.
It was pointed out by several authors that the application of an
appropriate design in these
studies can substantially increase the accuracy of estimation of the
population parameters.
Usually, the construction of a good design is based on the
Fisher information matrix which cannot be derived explicitly in
pharmacokinetic models with random effects.
For this reason, many authors propose an
approximation of the likelihood [see, e.g., \citet{menmalbac1997}, \citet{retmenbru2002}, \citet{retmen2003},
\citeauthor{schmelter2007a} (\citeyear{schmelter2007a,schmelter2007b}), among
others], which is used to derive an approximation for the Fisher
information matrix.
This matrix is considered in various optimality
criteria, which have been proposed for the construction of optimal
designs for random effect regression models.

In the present paper the investigation is motivated by the following issues.
First, the estimation of the population mean and the construction of
corresponding
optimal designs for population
pharmacokinetics strongly depends on the Gaussian assumption, which is
usually made for computational convenience. Moreover, the maximum likelihood
estimates may be inconsistent if the basic distributional assumption is
violated.
As a consequence, the derived optimal designs might be inefficient.
Second, most authors derive the approximation for the Fisher
information matrix under the additional assumption
that the random errors corresponding to the
measurements of each individual are uncorrelated [see, e.g., \citet
{redume2001}, \citet{retmenbru2002}
and \citet{retmen2003}, among many others].
However, this assumption is not realistic in many applications of
population pharmacokinetics.
Thus, a general concept for constructing optimal designs in the
presence of correlated observations is still missing.
Third, even if the Gaussian assumption
and the assumption of uncorrelated errors for each subject can be
justified, the numerical construction
of the estimate and the corresponding optimal design is extremely hard.

In the present paper we address these issues.
To derive a good design, we consider nonlinear least squares estimation
in random effect
regression models.
Note that this estimation does not require a specification of the
underlying distributions.
For this estimation, we introduce a methodology which can be used to
derive efficient or optimal designs in very general situations.
More precisely, we embed the discrete optimal design problem in a continuous
optimal design problem, where a nonlinear functional of the design
density has to be minimized or maximized.
This approach takes into account the correlation dependence
and yields an asymptotic optimal design density, which has to be
determined numerically in all cases of practical
interest.
For finding the optimal density, we propose an algorithm based on
polynomial approximation.
For a fixed sample size and for each individual,
an exact design can be obtained from the quantiles of the corresponding optimal
distribution function. It is demonstrated by examples that these
designs are very efficient.
Moreover,
the designs, derived from the asymptotic optimal design density, are
very good starting designs
for any procedure of local optimization for finding the exact optimal designs.
To our knowledge, the proposed method is the first systematic approach
to determine optimal designs
for linear and nonlinear mixed effect models with correlated errors.

The~remaining part of this paper is organized as follows. In
Section \ref{sec2} we consider the case of a linear random effect model and
explain the basic design concepts.
In Section \ref{sec3} we introduce the approach for deriving
the asymptotic optimal designs for linear regression models with
correlated observations
by employing results of \citet{bickherz1979}.
In Section \ref{sec4} we present results for nonlinear regression models with
correlated random errors
and derive $D$-optimal designs and optimal designs for estimating the
area under the curve
in the compartmental model.
Finally, the {Uzara\tsup{\textregistered}} and {Lanicor\tsup{\textregistered}}
trials are re-analyzed and
optimal designs for the model (\ref{3pramBat}) are determined.
In particular, we show that the design proposed by the experts is
rather efficient,
while a naively chosen equidistant design can yield substantially less
accurate estimates.

\section{Statement of the problem}\label{sec2}

Consider the common random effect linear regression model
%
\begin{eqnarray}\label{model-yij}
Y_{ij}=b_i^Tg(t_{ij})+\varepsilon_{ij}, \qquad i=1,\ldots, K;
j=1,\ldots,n_i,
\end{eqnarray}
where $Y_{ij}$ denotes the $j$th observation of the $i$th subject at
the experimental condition $t_{ij}$,
$\varepsilon_{11}, \ldots, \varepsilon_{K, n_K}$ are centered random
variables {with variances\vspace*{1pt} depending on~$t$,
$\operatorname{Var}(\varepsilon_{ij})=\sigma^2 h^2(t_{ij})$ for some
positive function $h(t)$},
$g(t)=(g_1(t),\ldots,g_p(t))^T$ is a given vector of linearly
independent regression functions, and $b_i$ is a $p$-dimensional random vector
representing the individual parameters of the $i$th subject,
$i=1,\ldots,K$.
The~explanatory variables $t_{ij}$ can be chosen by the
experimenter from a compact interval $T$. We assume that errors
$\varepsilon_{i} =
(\varepsilon_{i1}, \ldots, \varepsilon_{in_i})$ for different
subjects are independent, but the errors for the same subject are
correlated, that
is,
%
\begin{eqnarray}\label{corr}
\operatorname{Cov}(\varepsilon_{ij},\varepsilon_{is})=\sigma
^2h(t_{ij})h(t_{is})\bigl(\gamma r(t_{ij}-t_{is})+(1-\gamma)\delta
_{{j},{s}}\bigr),
\end{eqnarray}
where $\gamma\in[0,1]$ is a constant, $r(t)$ is a given correlation
function such that \mbox{$r (0)=1$}, and $\delta_{{j},{s}}$ denotes Kronecker's
symbol. Let $V_\varepsilon$ be the corresponding covariance matrix.
Assume that the individual parameters $b_i$ are drawn from a population
distribution
with mean $\beta$ and covariance matrix $V_p$, and
they are independent of the random variables $\varepsilon_{i}$.
This means that the covariance between two observations at the time
$t_{ij}$ and the time $t_{is}$ ($j \neq s$) is
\[
\operatorname{Cov} (Y_{ij}, Y_{is}) = g^T(t_{ij}) V_p g(t_{is}) + \sigma
^2h(t_{ij})h(t_{is}) \gamma r (t_{ij}-t_{is}) ,
\]
while the variance of $Y_{ij}$ is given by $ g^T(t_{ij}) V_p g(t_{ij})
+ \sigma^2 h^2(t_{ij}) $. It was shown by \citet{schmelter2007b} that
an optimal
design necessarily advises the experimenter to perform observations of
all subjects
at the same experimental settings, that is, $t_{ij} = t_j $
($i=1,\ldots
,K$, $j=1,\ldots, n$). Consequently, we define an exact design $ \xi
=\{t_1,\ldots,t_n\}$ as an $n$-dimensional
vector which describes the experimental conditions for each subject.
Without loss of generality, we assume that the design points are
ordered, $t_1<\cdots< t_n$.

Suppose that $n$ observations are taken according to the design $\xi$.
Then the model \eqref{model-yij} for the $i$th subject can be written as
%
\begin{eqnarray}\label{raneff}
{Y}_i=X_gb_i+{\varepsilon}_i;\qquad  i=1,\ldots,K,
\end{eqnarray}
where ${Y}_i=(Y_{i1},\ldots,Y_{in})^T$, the matrix $X_g$ is given by
$X_g=(g(t_1),\ldots,g(t_n))^T$, and ${\varepsilon}_i$ is now a
centered random variable with variance $\sigma^2h^2(t_i)$. This model
is a special case of the
random-effect models discussed in \citet{harville1976}, which are
called generalized MANOVA.
According to \citet{fedhack1997}, the (ordinary) least square estimate
of $\beta$ minimizes
\begin{eqnarray*}
\sum_{i=1}^K\sum_{j=1}^n h^{-2}(t_j)\bigl(Y_{ij}-f(t_j)\beta\bigr)^2.
\end{eqnarray*}
Define $f(t)=g(t)/h(t)$ and $X=(f(t_1),\ldots,f(t_n))^T$.
Then the covariance matrix\vspace*{1pt} of the ordinary least squares estimate $\hat
\beta_{\mathrm{OLS}}$
is given by
\begin{eqnarray}\label{covlse}
\mathbf{D}(\hat\beta_{\mathrm{OLS}})&=&\frac
{1}{K}(X^TX)^{-1}X^T(V_\varepsilon+XV_pX^T)X(X^TX)^{-1}\nonumber\\[-8pt]\\[-8pt]
&=&\frac{1}{K}\bigl((X^TX)^{-1}X^TV_\varepsilon
X(X^TX)^{-1}+V_p\bigr).\nonumber
\end{eqnarray}
Alternatively, if the covariance matrix
$V_\varepsilon$ of the errors and the covariance matrix of the random
effects $V_p$ were known
(or can be well estimated), the (weighted) least squares statistic
%
\begin{eqnarray}\label{WLS}
\quad \hat\beta_{\mathrm{WLS}}=\frac{1}{K}
\sum_{i=1}^K\bigl(X^T(V_\varepsilon+XV_pX^T)^{-1}X\bigr)^{-1}
X^T(V_\varepsilon+XV_pX^T)^{-1}Y_i,
\end{eqnarray}
can be used to estimate the parameter $\beta$.
The~covariance matrix of the estimate $\hat\beta_{\mathrm{WLS}}$ is
given by
%
\begin{eqnarray}\label{D-WLS}
\mathbf{D}(\hat\beta_{\mathrm{WLS}})=\frac{1}{K}
\bigl(X^T(V_\varepsilon+XV_pX^T)^{-1}X\bigr)^{-1}.
\end{eqnarray}

Since the expression \eqref{covlse} is simpler than \eqref{D-WLS}
(which requires
two different inversions),
the design methodology developed in this paper is based on the
covariance matrix
of the ordinary least squares estimate.
Additionally, we will demonstrate that the optimal designs obtained by
minimizing functionals of the covariance matrix
of the ordinary least squares estimate are also very efficient for
weighted least squares estimation.

We call a design optimal design if it minimizes an appropriate
functional of the covariance matrix
of the least squares estimate.
Since we consider optimality criteria that
are linear with respect to scalar multiplication of the covariance matrix,
we put $K=1$ without loss of generality.

\section{Asymptotic optimal designs}\label{sec3}

Although the theory of optimal design has been discussed intensively
for uncorrelated observations [see, e.g., \citet{fedorov1972},
\citet{pazman1986} and \citet{atkdon1992}], less results can be found
for dependent observations.
For linear and nonlinear random effect models, optimal designs under
the assumption of
uncorrelated errors have been investigated in
\citeauthor{schmelter2007a} (\citeyear{schmelter2007a,schmelter2007b}), \citet{menmalbac1997}
and \citet{retmen2003}, among others.
For fixed effect regression models with the presence of correlated
errors, it was suggested to derive optimal designs by asymptotic
considerations. \citeauthor{sackylv1966} (\citeyear{sackylv1966,sackylv1968}) have considered a
fixed design space, where the number of design points in this set tends
to infinity.
As a result, the asymptotic optimal
designs depend only on the behavior of the correlation function in a
neighborhood of the point $0$.
In the present paper we use the approach of
\citet{bickherz1979} and \citet{bickherzsch1981},
who have considered a design interval expanding proportionally to the
number of observation
points. This case is equivalent to the consideration of a fixed interval
with the correlation function depending on the sample size. To be precise,
we assume that the design space is given by an interval $T$ and
the design points of a sequence of designs
$\xi_n=\{t_{1n},\ldots,t_{nn}\}$ are generated by a function $a(\cdot
)$ in the form
%
\begin{eqnarray}
t_{jn}=a\bigl((j-1)/(n-1)\bigr),\qquad  j=1,\ldots,n;
\label{a-transf}
\end{eqnarray}
and $ a \dvtx  [0,1]\to T $ denotes the inverse of a distribution function.
Note that the function $a(\cdot)$ is obtained from the density of the
weak limit of the sequence $\xi_n$ as $n\to\infty$. For example, if
$T =[-1,1]$, the function $a(u)= (2u-1)$ corresponds to the equally-spaced
designs with distribution function $a^{-1} (x) = \frac{x+1}{2}$ and
density $(a^{-1})^\prime(x) = \frac{1}{2} I_{[-1,1]}(x)$.
Furthermore, we
assume that the correlation function $r(t)$
of the errors $\varepsilon_i$ in (\ref{corr})
depends on $n$ in the form
%
\begin{eqnarray}\label{rho0}
r_n(t)=\rho(nt),
\end{eqnarray}
such that $\rho(t)=o(t)$ as $t\to\infty$.
For the numerical construction of asymptotic optimal designs, we derive an
asymptotic representation for the covariance matrix of the least
squares estimate in the following lemma.
For this purpose, we impose the following regularity assumptions:
\begin{enumerate}[(C3)]
\item[(C1)]
The~functions $f_1(t),\ldots,f_p(t)$ are linearly independent, bounded
on the interval $T$
and satisfy the first order Lipschitz condition:
\begin{eqnarray*}
|f_i(t)-f_i(s)|&\le& M|t-s| \quad \mbox{and}\\
  |f_i(t)|&\le& M
\qquad \mbox{for all } t,s\in T, i=1,\ldots,p.
\end{eqnarray*}
\item[(C2)]
The~function $a(\cdot)$ is twice differentiable and there exists a
positive constant $M<\infty$ such that for all $ u\in(0,1)$,
%
\begin{eqnarray}
\label{bound}
\frac{1}{M}\le a'(u) \le M,\qquad  |a''(u)|\le M.
\end{eqnarray}
\item[(C3)]
The~correlation function $\rho$ is differentiable with bounded
derivative and satisfies
$\rho'(t)\le0$ for sufficiently large $t$.
\end{enumerate}
Assumption (C1) refers to the continuity of the response as a function
of $t$ and
it is satisfied for most of the commonly used regression models.
Moreover, most of the commonly used
correlation structures satisfy assumption (C3) on a compact interval.
Finally, assumption (C2)
refers to the characteristics of a design, requiring, loosely speaking,
that observations should not be clustered. This is quite reasonable due
to ethical aspects.
The~following result is obtained by similar arguments as given in \citet
{bickherz1979}
and, hence, its proof is omitted.

\begin{lemma*}
Assuming that conditions \textup{(C1)}, \textup{(C2)} and \textup{(C3)} are
satisfied, then the covariance matrix of the
least squares estimate have the form
%
\begin{eqnarray}\label{e5}
\quad \mathbf{D} (\hat\beta_{\mathrm{OLS}}) = \frac{\sigma^2}{n}
\bigl(W^{-1}(a)+2\gamma
W^{-1}(a)R(a)W^{-1}(a)\bigr)+V_p +o\biggl(\frac{1}{n}\biggr),
\end{eqnarray}
where the matrices $W(a)$ and $R(a)$ are defined by
\begin{eqnarray*}
W(a)&=&\biggl(\int_0^1 f_i(a(u)) f_j(a(u))  \,du\biggr)_{i,j=1}^p, \\
R(a)&=&\biggl(\int_0^1 f_i(a(u)) f_j(a(u)) Q(a'(u))\, du
\biggr)_{i,j=1}^p
\end{eqnarray*}
and the function $Q(\cdot)$ is given by
$
Q(t)=\sum_{j=1}^\infty\rho(jt).
$
\end{lemma*}

Note that only the first term in (\ref{covlse}) [and (\ref{e5})]
depends on a (asymptotic) design,
and we propose to use this term
for the construction of optimal designs.
If the function $a(\cdot)$ is the inverse of a continuous distribution
with density, say, $\varphi$,
then $a'(t)=1/\varphi(t)$ and, for large $n$, the first term of the
covariance matrix can be approximated by the matrix $V(\varphi)/n$,
where the $p \times p$
matrix $V$ is given by
%
\begin{eqnarray}\label{inf} V(\varphi)
&:=&\sigma^2\bigl(W^{-1}(\varphi)+2\gamma W^{-1}(\varphi)R(\varphi
)W^{-1}(\varphi)\bigr).
\end{eqnarray}
In \eqref{inf} the matrices $W(\varphi)$ and $R(\varphi)$ have the form
\begin{eqnarray*}
W(\varphi)&=&\biggl(\int_{T} f_i(t) f_j(t)\varphi(t)\,  dt
\biggr)_{i,j=1}^p, \\
R(\varphi)&=&\biggl(\int_{T} f_i(t) f_j(t) Q\bigl(1/\varphi(t)\bigr)\varphi
(t) \,dt\biggr)_{i,j=1}^p.
\end{eqnarray*}
A density will be called an asymptotic optimal density $\varphi^*$, if it
minimizes an appropriate functional of the matrix $V(\varphi)$.
Numerous criteria have been proposed in
[\citet{silvey1980}, \citet{atkdon1992}, \citet{pukelsheim1993}] and,
exemplarily, we consider the $D$- and $c$-optimality criteria which minimize
$-\det V(\varphi)$ and $c^TV(\varphi) c$ for a given vector $c \in
\mathbb{R}^p$, respectively.
The~application of the proposed methodology to other
optimality criteria is straightforward.
The~general procedure for constructing an efficient design minimizing a
given functional of the covariance matrix
of the least squares estimate is as follows:
\begin{enumerate}[(3)]
\item[(1)] Specify the correlation function $\rho(\cdot)$ in (\ref
{rho0}) and compute $Q(\cdot)$.
\item[(2)] Compute the asymptotic optimal design density $\varphi^*$
that minimizes
an appropriate functional of the matrix $V(\varphi)$ in (\ref{inf}).
\item[(3)] Derive the exact design for a fixed sample
size $n$ by calculation of the quantiles of the distribution function
$\Phi^*$ that corresponds to $\varphi^*$,
namely,
%
\begin{eqnarray}\label{trafo}
t_{i,n} = (\Phi^*)^{-1}\biggl({i-1\over n-1}\biggr);\qquad  i=1,\ldots,n.
\end{eqnarray}
\end{enumerate}
The~optimal density $\varphi^*$ in step (2) can be determined as follows.
We consider the parametric representation of a
density by a polynomial in the form
\begin{eqnarray*}
\varphi(t)=\frac{(p_0+p_1t+\cdots+p_rt^r)_+}{\int_{T}
(p_0+p_1t+\cdots+p_rt^r)_+ \,dt}
\end{eqnarray*}
and apply the Nelder--Mead algorithm to find the optimal density
that minimizes the specified functional of the matrix $V(\varphi)$
with respect to $p_0,\ldots,p_r$.
One can run the algorithm for different degrees of the polynomial and
different initial values and choose the density
corresponding to the minimal value of the optimality criterion.
All integrals can be calculated by the Simpson quadrature formula.
We found that the minimal value of the criterion is negligibly
decreasing for
polynomials of degree larger than $r=6$.
We also investigated the case where a density is
represented in terms of rational, exponential or spline functions.
The~results were very similar and, on the basis of our numerical experiments,
we can conclude that the optimal density $\varphi^*$ can be very well
approximated by 6-degree polynomials.

The~derived designs from the asymptotic theory
can be used to construct efficient designs for a given sample size as
specified in step (3) of the algorithm.
Alternatively, exact optimal designs can be determined by employing the
derived designs
as initial values in a (discrete) optimization procedure.
More precisely, for the determination of an exact optimal design,
the above procedure can be extended by the fourth step:
\begin{enumerate}[(4)]
\item[(4)] Determine an exact optimal design, that minimizes a
functional of the covariance matrix in (\ref{covlse}),
by using the Nelder--Mead algorithm with an initial $n$-point design,
which has been found in step~(3).
\end{enumerate}
We illustrate this methodology for a quadratic regression model by the
following example.
Also, in Section \ref{sec4} we extend the approach for designing for nonlinear
models and
investigate its performance for the compartmental model.
In particular, we demonstrate that the designs derived from the
asymptotic consideration
are very efficient compared to exact optimal designs.

\newcommand{\myw}{}

\begin{figure}

\includegraphics{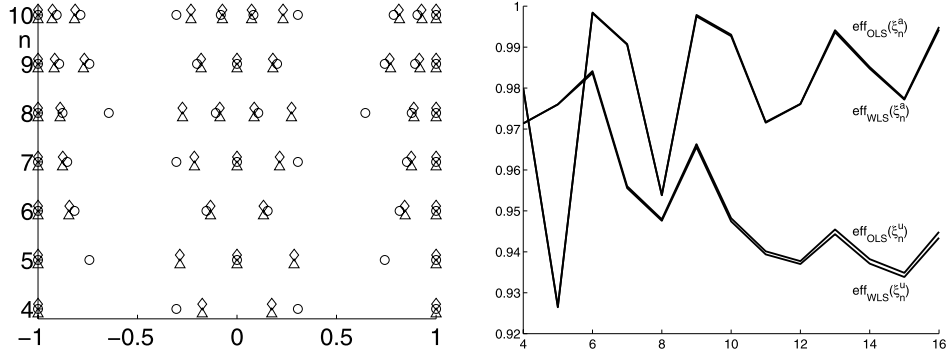}

\caption{Left part: Various designs for ordinary and weighted
least squares estimation in the random effect quadratic
regression model.
Exact $D$-optimal designs derived from asymptotic theory: ball;
exact $D$-optimal for ordinary least squares estimation: diamond;
exact $D$-optimal designs for weighted least squares estimation: triangle.
Right part: Efficiency of the designs $\xi_n^a$ and equidistant design
$\xi_n^u$ for ordinary and weighted least squares estimation.
The~parameters are given by \textup{(\protect\ref{rho1})}, where $ \lambda=1.2$.}
\label{fig:2comp2}
\end{figure}

\begin{example}[(Quadratic regression)]\label{ex1}
We illustrate the proposed approach for constructing optimal designs
for the classical quadratic regression model with homoscedastic errors.
For this model, we have $p=3$, $f_1(t)=1$, $f_2(t)=t$, $f_2(t)=t^2$ and
$T=[-1,1]$.
Let the correlation
function in (\ref{rho0}) be given by $\rho(t) = e^{- \lambda t}$ and
%
\begin{equation}\label{rho}
r_n (t) = e^{-\lambda nt},\qquad  \lambda>0.
\end{equation}
The~asymptotic $D$-optimal densities for different choices of the parameters
$\lambda$ and $\gamma$ are shown in Figure~\ref{fig:qdens1}.
Note that the numerically calculated optimal design densities are
symmetrical, but we were not able to prove the symmetry of the
asymptotic optimal density.
We observe that the $D$-optimal density converges to the density of the
uniform design, if $\gamma\to1$ or $\lambda\to0$. On the other
hand, if
$\gamma\to0$ or $\lambda\to\infty$, it can be seen that the
asymptotic $D$-optimal design density is more concentrated at the
points $-1$, $0$
and $1$, which are the points of the exact $D$-optimal design for the
quadratic fixed effect model with uncorrelated observations
[see \citet{gaffkra1982}]. Such behavior is natural because the errors
are less correlated as
$\gamma\to0$ or $\lambda\to\infty$.

\begin{figure}

\includegraphics{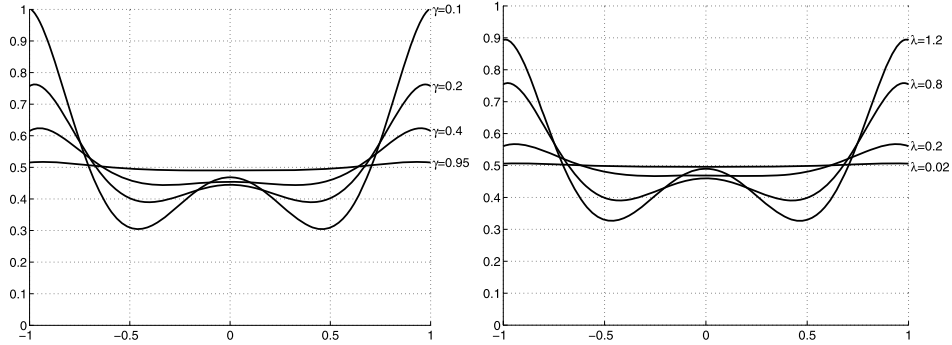}

\caption{Asymptotic $D$-optimal design densities for least squares
estimation in a random effect quadratic model for different
choices of parameters in the covariance function
\textup{(\protect\ref{corr})} with $r(t)$ defined by \textup{(\protect\ref{rho})}. Right panel $\gamma
= 0.6$; left panel $\lambda= 0.2$.}
\label{fig:qdens1}
\end{figure}

Further, we investigate the efficiency of an exact design derived from
the asymptotic theory
for least squares estimation, where the parameters in the
correlation function \eqref{corr} are given by
%
\begin{equation} \label{rho1}
\quad \gamma=0.6,\qquad  \sigma^2=0.5, \qquad  V_p=\operatorname{diag}(\sigma_{\beta
_1}^2,\sigma_{\beta_2}^2) =\operatorname{diag}(0.3^2,0.3^2).
\end{equation}
Let $\xi_n^u$ be an $n$-point equidistant design and $\xi_n^a$ be an
$n$-point design obtained by the transformation (\ref{trafo})
from the asymptotic optimal density.
The~points of the design $\xi_n^a$ and the exact optimal designs for
least squares estimation for $\lambda=1.2$ are displayed in the left
part of Figure \ref{fig:2comp2}.
In the right part of Figure \ref{fig:2comp2}, we present the efficiencies
for ordinary least squares estimation,
\begin{eqnarray*}
\operatorname{eff}_{\mathrm{OLS}} (\xi) &=& \biggl( { \det[
(X^TX)^{-1}X^TV_\varepsilon X(X^TX)^{-1}+V_p ] \over\det[
(X^T_{\mathrm{OLS}}X_{\mathrm{OLS}})^{-1}X^T_{\mathrm
{OLS}}V_\varepsilon
X_{\mathrm{OLS}}(X^T_{\mathrm{OLS}}X_{\mathrm{OLS}})^{-1}+V_p ]
}\biggr)^{1/p},
\end{eqnarray*}
and for weighted least squares estimation,
\begin{eqnarray*}
\operatorname{eff}_{\mathrm{WLS}} (\xi) &=& \biggl( { \det[
X^T(V_\varepsilon+XV_pX^T)^{-1}X ] \over\det[
X^T_{\mathrm{WLS}}(V_\varepsilon+X_{\mathrm{WLS}}V_pX^T_{\mathrm
{WLS}})^{-1}X_{\mathrm{WLS}} ] }\biggr)^{1/p}
\end{eqnarray*}
for two designs:
the design $\xi_n^a$, derived from asymptotic theory,
and the uniform design $\xi_n^u$.
Here $X$ denotes the design matrix obtained for the design $\xi$ under
investigation,
while $X_{\mathrm{OLS}}$ and $X_{\mathrm{WLS}}$ correspond to the
optimal exact design
for ordinary and weighted least squares estimation respectively.
We observe that the design
points concentrate in three regions containing the points of the exact
\mbox{$D$-optimal} design for a quadratic regression with uncorrelated errors. It
is also noteworthy that the designs derived from the asymptotic theory
are very efficient, in particular, for weighted least squares estimation.
\end{example}

\section{Nonlinear random effect models}\label{sec4}

In this section we extend the methodology to the case of nonlinear
random effect models,
which have found considerable interest in the literature on pharmacokinetics.
In this case, the results of experiments are modeled by
%
\begin{eqnarray}\label{nonlinear}
Y_{ij}= \eta( t_{j},b_i)+\varepsilon_{ij},\qquad  i=1,\ldots, K;
j=1,\ldots, n.
\end{eqnarray}
Since the model \eqref{nonlinear} is nonlinear with respect to the
variables $b_i$,
there is no analytical expression for the
likelihood function and various approximations have been considered in
the literature [see \citet{menmalbac1997}, \citet{retmenbru2002}, \citet
{retmen2003}, among others].
These approximations are used for the calculation of the maximum
likelihood estimate and
the corresponding Fisher information matrix. Alternatively,
an estimate of the population mean $\beta$ can be obtained as an
average of the nonlinear least squares estimates $\hat b_i$ for the different
individuals, but due to the nonlinearity of the model, an explicit
representation of the corresponding covariance matrix cannot be derived.
Following \citet{retmen2003}, we employ a first-order Taylor expansion
to derive an approximation
of this covariance matrix. To be precise, we obtain (under suitable
assumptions of differentiability
of the regression function) the approximation
%
\begin{eqnarray}\label{linear}
\eta(t,b)\approx\eta(t,\beta)+g(t,\beta)(b-\beta)^T,
\end{eqnarray}
where
\begin{eqnarray*}
g(t,b)=\frac{\partial\eta(t,b)}{\partial b}
\end{eqnarray*}
denotes the gradient of the regression function with respect to $b$.
This means that the nonlinear model \eqref{nonlinear} is approximated
by the linear model \eqref{linear}.
For the construction of the optimal design,
we assume that knowledge about the parameter $\beta$ is available from
previous or similar experiments.
This corresponds to the concept
of locally optimal designs, introduced by \citet{chernoff1953} in the
context of fixed effect nonlinear regression models.
Usually, locally optimal designs serve as benchmarks for commonly used
designs and are the basis for the construction of
optimal designs with respect to more sophisticated optimality criteria including
the Bayesian and minimax approach [see \citet{chaver1995} or \citet
{dette1995}].

Following the discussion in Section \ref{sec3}, we define the function
$f(t,b)=g(t,b)/h(t)$ to account for heteroscedasticity.
Note that
the covariance matrix of the nonlinear least squares estimate
in the model \eqref{nonlinear} is approximated by
replacing the matrix $X$ in model \eqref{model-yij} with
$f(t)=f(t,b)|_{b=\beta}$, and the methodology described
in Sections \ref{sec2} and \ref{sec3} can be applied to
determine efficient designs. In the context of dose finding studies, it
has been shown by means of
a simulation study that the approximation (\ref{linear}) has
sufficient accuracy for the construction
of optimal designs [see Section 5 in \citet{debrpepi2008} for more details].

We further illustrate this concept by giving several examples for the
case of homoscedastic errors.
First, we investigate $D$- and $c$-optimal designs for the random
effect model,
which has recently been studied by \citet{atkinson2008}.
Next, we re-analyze the {Uzara\tsup{\textregistered}} and
{Lanicor\tsup{\textregistered}} trials introduced in Section~\ref{sec1}.

\subsection{$D$-optimal design for a random effect compartmental model}\label{sec4.1}

We consider the random effect compartmental model with first-order absorption,
%
\begin{eqnarray}\label{comp.model}
\eta(t,b)=\frac{b_1}{b_1-b_2}(e^{-b_2 t}-e^{-b_1 t}).
\end{eqnarray}
The~model (\ref{comp.model}) is a special case of the Bateman
function, defined
in the introduction [see \citet{garrett1994}],
and has found considerable attention in chemical sciences, toxicology and
pharmacokinetics [see, e.g., \citet{gibper1982}].
The~optimal design problem in the compartmental fixed effect model has
been studied by
numerous authors [see, e.g.,
\citet{boxluc1959}, \citet{atchheju1993}, \citet{detobr1999}, \citet
{biedetpep2004}, among others],
but much fewer results are available under the assumption of random effects.
Recently optimal approximate designs for a random-effect compartmental model
\eqref{comp.model} have been determined by \citet{atkinson2008}, but
we did not find results about exact designs for these models in the
presence of correlated errors.
In the present paper we derive such designs from the asymptotic optimal
design density and
compare these designs with the exact optimal designs.

\begin{figure}[t]

\includegraphics{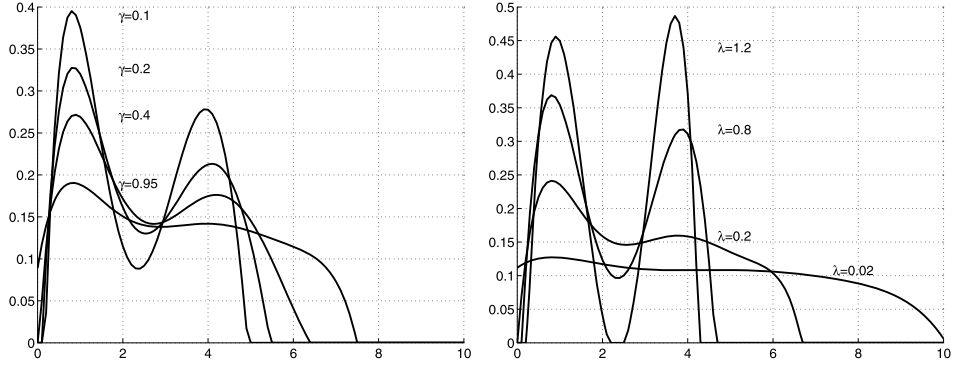}

\caption{Asymptotic $D$-optimal design densities
for nonlinear least squares estimation in the compartmental model
\textup{\protect\eqref{comp.model}}
for different choices of the parameters in the covariance function
\textup{(\protect\ref{corr})} with $r(t)$ defined by \textup{(\protect\ref{rho})}. Left part:
$\lambda=0.2$; right part: $\gamma=0.6$.}
\label{fig:densn}\vspace*{-4,8pt}
\end{figure}

Note that the gradient of the function $\eta(t,b)$ with respect to $b$
is given by
\begin{eqnarray}\label{nonl-f}
g(t,b)&=&\biggl(\frac{b_2({e^{-b_1x}}-{e^{-b_2x}})+
({b_1}^{2}x-b_1b_2x){e^{-b_1x}}}
{ ( b_1-b_2 ) ^{2}},\nonumber\\[-8pt]\\[-8pt]
&&\hspace*{11pt}
\frac{b_1({e^{-b_1x}}-{e^{-b_2x}})+
({b_1}^{2}x-b_1b_2x){e^{-b_2x}}}
{ ( b_1-b_2 ) ^{2}}\biggr)^T.\nonumber
\end{eqnarray}
In order to illustrate the methodology, we consider the same scenario
as in \citet{atkinson2008} and
assume that the parameters of the population and error distributions
are the following:
$\beta^{(0)}=(1,0.5)^T$,
%
\begin{equation} \label{rho2}
\qquad \gamma=0.6,\qquad  \sigma^2=0.01,\qquad  V_p=\operatorname{diag}(\sigma_{\beta
_1}^2,\sigma_{\beta_2}^2) =\operatorname{diag}(0.1^2,0.05^2),
\end{equation}
and the design space is given by the interval $T= [0,10]$.
We assume that the function $r(t)$ in (\ref{corr}) is given by (\ref{rho}).
The~asymptotic $D$-optimal design densities for different values of the
parameters are shown in Figure \ref{fig:densn}. We observe that, for
$\gamma\to1$ or $\lambda\to0$,
the $D$-optimal design densities approximate the uniform design,
while, for larger values of $\lambda$ or small values of $\gamma$,
the asymptotic $D$-optimal designs put more weight at two specific
regions of the design space.
This fact corresponds to intuition, because the (approximate)
$D$-optimal design
for the model (\ref{comp.model}) with uncorrelated
observations is a two-point design [see, e.g., \citet{boxluc1959}].

\begin{figure}[b]\vspace*{-4,8pt}

\includegraphics{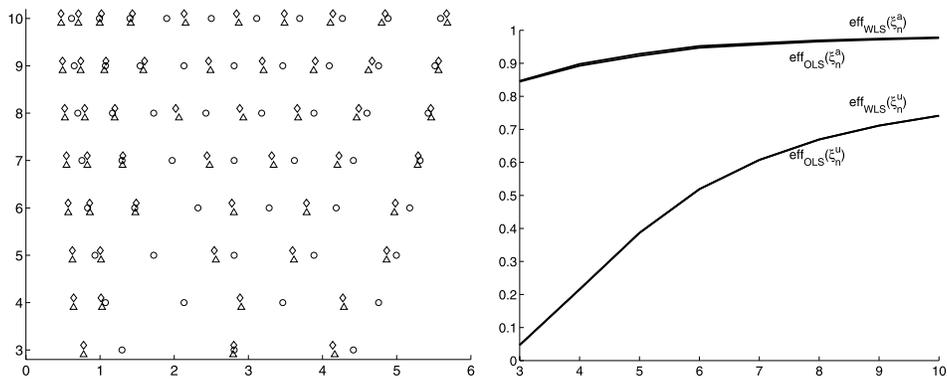}

\caption{Left part: Various designs for ordinary and weighted
nonlinear least squares estimation in
the compartmental model \textup{\protect\eqref{comp.model}}.
Exact $D$-optimal designs derived from asymptotic theory: ball;
exact $D$-optimal for ordinary least squares estimation: diamond;
exact $D$-optimal designs for weighted least squares estimation: triangle.
Right part: Efficiency of the designs $\xi_n^a$ and $\xi_n^u$ for
ordinary and weighted least squares estimation.
The~parameters are given by \textup{(\protect\ref{rho2})}, where $ \lambda=0.2$.}
\label{fig:compn}\vspace*{-2,5pt}
\end{figure}

Further, we investigate the performance of the uniform and an exact
design derived from asymptotic theory. For this purpose we define
$\xi_n^u$ as an $n$-point equidistant design $\{10/n,20/n,\ldots,10\}$
and $\xi_n^a$ as the $n$-point design obtained by the transformation
\begin{eqnarray*}
t_{j}=(\Phi^*)^{-1}\bigl(j/(n+1)\bigr),\qquad  j=1,\ldots,n,
\end{eqnarray*}
where $\Phi^*$ denotes the distribution function corresponding to
the asymptotic \mbox{$D$-optimal} design density.
Note this transformation is slightly different from the transformation
\eqref{a-transf} in
order to exclude the point $0$ from the design points. Obviously, it is
not reasonable to take observations $t=0$ in model
\eqref{comp.model}, because it is assumed that the drug is
administered at time $t=0$. The~corresponding points of the exact
designs are depicted in the left part of Figure \ref{fig:compn}, while
the right part
of the figure shows the $D$-efficiencies of the different designs.
We observe that the designs derived from the asymptotic theory have a
substantially larger $D$-efficiency compared to the uniform design.
For example, for $n=6$, the $D$-efficiency of the uniform design is
approximately $50\%$ {for ordinary and weighted nonlinear least squares
estimation}, while the $D$-efficiency of the design $\xi_n^a$ is close to
$90\%$.

\begin{figure}

\includegraphics{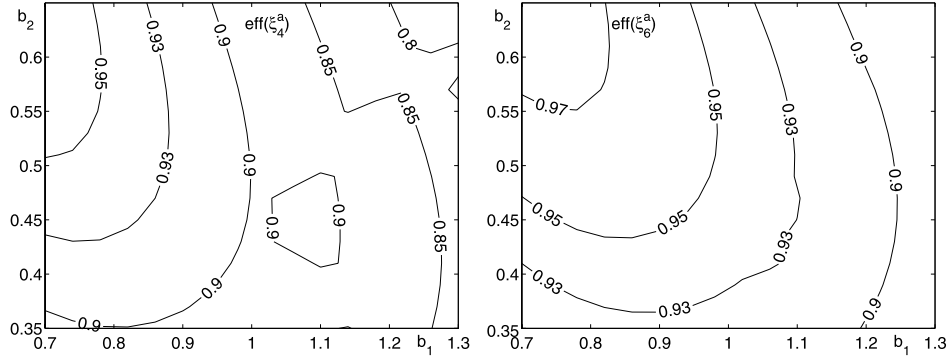}

\caption{$D$-efficiencies of the designs $\xi^a_4=\{1.04, 2.01,
3.16, 4.33\}$ (left part)
and the design $\xi^a_6=\{0.83, 1.47, 2.32, 3.30, 4.20, 5.20\}$
(right part) (these designs are obtained from asymptotic density) for
nonlinear least squares estimation in the {random effect} compartmental
model \textup{\protect\eqref{comp.model}}, if the mean of the population
distribution has been misspecified. The~parameters of the population
distribution are given by \textup{\protect\eqref{rho2}} with $\lambda=0.2$.}
\label{fig:eff46a}\vspace*{-2pt}
\end{figure}

It is worthwhile to mention that, in nonlinear random effect models,
the optimal designs depend additionally on the mean $\beta$ of the
distribution of the population parameters
$b_i$. Therefore, it is of interest to investigate the sensitivity of
the designs with respect to a misspecification of this parameter. For
a study of the impact of such a misspecification on the efficiency of
the resulting designs,
we consider the case $n=4$ and $n=6$ and the
corresponding designs $\xi^a_4=\{1.04, 2.01, 3.16, 4.33\}$ and $\xi
^a_6=\{0.83, 1.47, 2.32, 3.30, 4.20, 5.20\}$, respectively. In Figure
\ref{fig:eff46a} we display the efficiencies
%
\begin{eqnarray}
\label{eff1}
\hspace*{19pt}\biggl( { \det[ (X^TX)^{-1}X^TV_\varepsilon
X(X^TX)^{-1}+V_p ] \over\det[
(X^T_{\mathrm{OLS},\beta}X_{\mathrm{OLS},\beta})^{-1}X^T_{\mathrm
{OLS},\beta}V_\varepsilon
X_{\mathrm{OLS},\beta}(X^T_{\mathrm{OLS},\beta}X_{\mathrm
{OLS},\beta})^{-1}+V_p ] }\biggr)^{-1/p}\hspace*{-19pt}
\end{eqnarray}
for different values of $\beta$. Here $X$
denotes the design matrix obtained from the design $\xi_n^a$ under the
assumption that $\beta^{(0)} = (1,0,5)^T$, while the matrix
$X_{\mathrm{OLS},\beta}$
corresponds to the exact $D$-optimal design for nonlinear least squares
estimation for a specific $\beta$. The~efficiencies are plotted in
Figure \ref{fig:eff46a} for the rectangle
$[\beta_i^{(0)}-3\sigma_{\beta_i^{(0)}},\beta_i^{(0)}+3\sigma
_{\beta_i^{(0)}}] = [0.7, 1.3] \times[0.35, 0.65]$.
It can be seen that the exact optimal design, derived from the
asymptotic theory,
has a high $D$-efficiency with respect to misspecification of the
population mean $\beta$
over a broad range.

%

%

\subsection{Optimal designs for estimating the AUC}\label{sec4.2}

\begin{figure}

\includegraphics{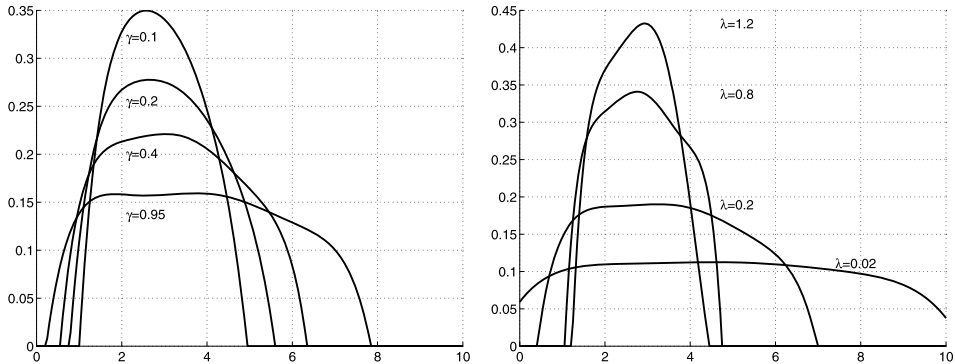}

\caption{Asymptotic optimal densities for estimating the area
under the curve in the compartmental model \textup{\protect\eqref{comp.model}} for different
choices of the parameters in the covariance function
\textup{(\protect\ref{corr})} with $r(t)$ defined by \textup{(\protect\ref{rho})}. Left part: $\lambda
=0.2$, right part: $\gamma=0.6$.}
\label{fig:densn_auc}
\end{figure}

\begin{figure}

\includegraphics{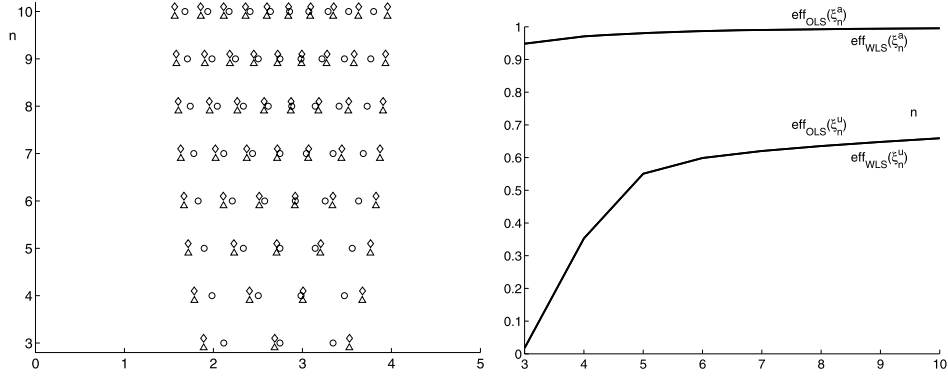}

\caption{Left part: Various designs for estimating the area under
the curve in the compartmental model
\textup{\protect\eqref{comp.model}}.
Designs derived from asymptotic theory: ball;
exact optimal designs for least squares estimation of the area under
the curve: diamond;
exact optimal designs for weighted least squares estimation: triangle.
Right part: Efficiency of the designs $\xi_n^a$ and $\xi_n^u$ for
ordinary and
weighted least squares estimation of the AUC. The~parameters are given
by \textup{\protect\eqref{rho2}}, where $ \lambda=1.2$.}
\label{fig:compn2_auc}
\end{figure}

In some bioavailability studies, the aim of experiments is the
estimation of the area under curve
\begin{eqnarray*}
\mathrm{AUC}=\int_0^\infty\eta(t,\beta)\, dt.
\end{eqnarray*}
For the
compartmental model \eqref{comp.model}, we obtain $\mathrm{AUC}={1}/{b_2}$.
It can be shown that the locally AUC-optimal design for model \eqref
{comp.model} minimizes
the variance of the nonlinear least squares estimate for the parameter
$\beta_2$.
This variance is approximately proportional to
\[
(0,1) \bigl((X^TX)^{-1}X^TV_\varepsilon X(X^TX)^{-1}+V_p\bigr) (0,1)^T.
\]
This expression corresponds to the $c$-optimality criterion,
which has been discussed extensively in the literature for fixed effect
models with uncorrelated
observations [see, e.g., \citet{fortorwu1992}, \citet
{fanchal2003} and \citet{debrpepi2008}, among others].
The~asymptotic optimal design densities for estimating the area under
the curve are shown in Figure \ref{fig:densn_auc}. We observe
again that the optimal density is close to the uniform design density
if $\lambda\to0$ or $\gamma\to1$. On the other hand, if $\lambda$
is large or
$\gamma\to0$, the AUC-optimal design density has a narrow support.
This fact reflects that the optimal design for estimating the area
under the curve in the fixed effect compartmental model with
uncorrelated observations is a one-point design.
In Figure \ref{fig:compn2_auc} we show the designs derived from the
asymptotic optimal design densities,
and the exact optimal designs for estimating the
area under the curve in the compartmental model.
We observe that the designs $\xi^a_n$, derived from the asymptotic
optimal design density, are very close
to the exact optimal designs for least squares estimation of the area
under the curve. Moreover, the design $\xi_n^a$ yields a
substantial improvement in efficiency compared to the uniform design.
Again we observe that the designs derived for ordinary least
squares estimation also have excellent efficiencies for weighted least
squares estimation of the AUC (see the right part of Figure
\ref{fig:compn2_auc}).

~~~~~~~

\subsection[Optimal designs for estimating the AUC in the
{Uzara} and {Lanicor} trials]{Optimal
designs for estimating the AUC in the
{Uzara\tsup{\textregistered}} and {Lanicor\tsup{\textregistered}} trials}

In this section we consider the optimal
design problem for estimating the AUC in the two
examples presented in the introduction. The~medical background of these
examples was a small pilot trial of 4 patients [\citet{thuermann2004}],
where it was observed that patients taking the over-the-counter herbal
diarrhea medication Uzara\tsup{\textregistered}~(in the form of drops,
i.e., oral application) showed high values in medical assays designed
to measure the blood serum concentration of digitoxin, a potent
treatment against heart insufficiency. This is caused by the chemical
similarity of these two substances and can result in major
complications in establishing treatment programs for heart
insufficiency. It was thus decided to compare the pharmacokinetic
properties of an oral application of Uzara\tsup{\textregistered}~to the
properties of the usual intravenous application of a regular digitoxin
medication (Lanicor\tsup{\textregistered}) on a larger sample size of 18
patients, with 15 measurements each. The~main focus of the comparison
was the area under the concentration curve as a measure of the total
effect of an application. A preliminary design was proposed by experts
in order to allow precise estimation of this property, and the study
was carried out according to this design. We will now investigate
the efficiency of this and a naively chosen equally spaced design
compared to a design generated using the methods presented in this
paper.

In order to compute the design densities, it is necessary to estimate
the parameters for both of the models.
{To do so, we used the full 18 patient data set, however, this could
have been done as well using only the 4 patients of the pilot study.}
We estimated parameters using a combination of maximum likelihood and
least squares techniques [see \citet{pinbat1995}]. For the intravenous
part corresponding to model \eqref{2par},
we have received the estimates $ \hat\beta=(30,0.75)^T$,
\[
\hat
V_p=
\pmatrix{ 9&0.189 \vspace*{2pt}\cr
0.189&0.0049
}
\]
for the population parameters, while the estimates of the parameters in the
covariance matrix \eqref{covpop} are given by $ \hat
\gamma=0.8$, $\hat\lambda=0.05$, $\hat\sigma^2=0.6$.

For the oral application corresponding to model \eqref{3pramBat},
the estimates of the parameters are given by $ \hat\beta=(0.2,0.135,28)^T$,
\[
\hat
V_p=
\pmatrix{ 0.0025&0.0019 &0\vspace*{2pt}\cr
0.0019&0.0016&0\vspace*{2pt}\cr
0&0&144}
\]
and $ \hat
\gamma=0.8$, $\hat\lambda=0.01$, $\hat\sigma^2=0.2$. Note that the
entries in the positions $(1,3)$, $(2,3)$, $(3,1)$ and $(3,2)$
are not identical $0$, but smaller than $10^{-5}$.
Based on a preliminary discussion with experts,
the physicians decided
to take measurements for both experiments at nearly identical
(nonoptimized) time points
\[
0, 0.5, 1,1.5, 2, 3,
4, 5, 6, 8, 10, 12, 15, 24, 36
\]
and
\[
0.25, 0.75,1.25, 2, 3,
4, 5, 6, 8, 10, 12, 15, 24, 36,
\]
respectively. 
Note that in the second design an additional measurement was taken at $t=-0.25$,
that is, before the intravenous injection. Since this point is out of
the scope of the exponential evasion model,
the observation at this time has been excluded from further considerations.
Thus, the {Lanicor\tsup{\textregistered}} trial has $n=14$ observations.

\begin{figure}[b]

\includegraphics{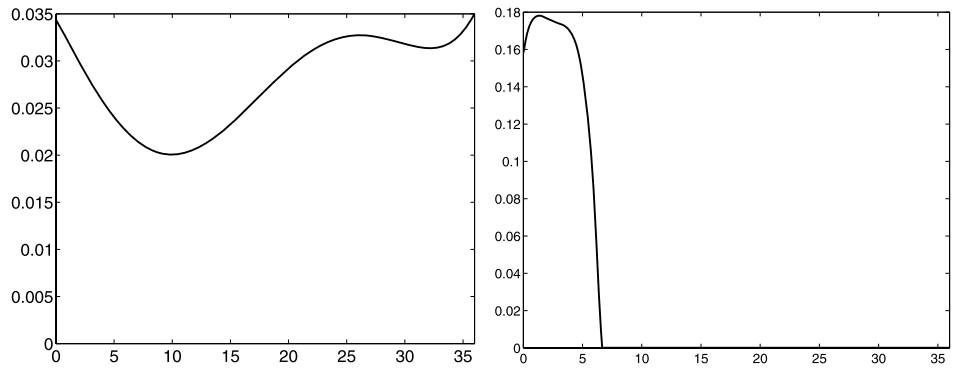}

\caption{Asymptotic optimal density for estimating the area under
the curve in the $3$-parameter Bateman
(left part) and exponential evasion model (right part).}
\label{fig:uzara_densn_auc}
\end{figure}

For the estimated parameters, we have derived the asymptotic optimal
design densities, which
are depicted in the left and right part of Figure \ref{fig:uzara_densn_auc}
for
the {Uzara\tsup{\textregistered}} and {Lanicor\tsup{\textregistered}} trials,
respectively, where
the design interval is given by $T= [0,36]$.
The~resulting designs from these densities are given by
\[
2.09,  4.55,  7.49,  10.8,  13.9,  16.8,  19.2,  21.5,  23.6,  25.7,
27.7,  29.8,  31.9,  34.0,  36
\]
for the {Uzara\tsup{\textregistered}} trial and
\[
0.432,  0.85,  1.25,  1.66,  2.06,  2.46,  2.87,  3.28,  3.7,  4.12,
4.55,  5.01,  5.54,  36
\]
for the {Lanicor\tsup{\textregistered}} trial. This means that the asymptotic
optimal design for the oral application
is close to an equally spaced design, while the optimal design for the
intravenous application is much more focused on
measurements at early time points. This result is plausible in an
exponential elimination model.
Calculating the efficiencies for ordinary least squares estimation of
these asymptotic optimal designs compared to the exact optimal design,
we obtain efficiencies $0.99$ and $0.95$, respectively, indicating that
these designs are quite good.

Upon numerical studies, we can further conclude that the constructed
designs are robust to
misspecification of the initial guess of parameters.
For example, variations of $\lambda$ (or any of the other parameters)
by up to 50\% yield a drop of efficiency
by less than $0.02$.

Note that in the examples presented in Sections \ref{sec3}, \ref{sec4.1} and \ref{sec4.2}, the
efficiencies of the derived designs
for weighted least squares estimation are very similar to the
efficiencies for ordinary least squares estimation, and a similar
observation has been made for the two trials under investigation. The~efficiencies for weighted least squares estimation of
the designs based on the asymptotic density are 99\%
({Uzara\tsup{\textregistered}}) and 97\% ({Lanicor\tsup{\textregistered}}),
even better than the efficiencies for ordinary least squares estimation.

We now investigate the performance of the designs which were actually
used in the clinical trial.
We found that these designs have efficiency $0.96$ and $0.92$
for estimating the AUC through OLS estimation in the
{Uzara\tsup{\textregistered}} and {Lanicor\tsup{\textregistered}} trials, respectively.
Thus, the preliminary designs, recommended by experts, are rather efficient
in both trials.

Let us now investigate the performance of
naively chosen equidistant designs: the design
\[
0,  2.6,  5.1,  7.7,  10.3,  12.8,  15.4,  18.0,  20.6,  23.1,  25.7,
28.3,  30.8,  33.4,  36
\]
for the {Uzara\tsup{\textregistered}} trial, and the design
\[
0,  2.8,  5.5,  8.3,  11.1,  13.9,  16.6,  19.4,  22.2,  24.9,  27.7,
30.5,  33.2,  36
\]
for the {Lanicor\tsup{\textregistered}} trial.
Comparing these designs to the optimal designs, we obtain efficiencies
of $0.97$ and $0.41$
for ordinary least squares estimation of the AUC in
the {Uzara\tsup{\textregistered}} and {Lanicor\tsup{\textregistered}} trials,
respectively.

It was pointed out by a referee that it is of some interest to
investigate the performance of the
optimal designs proposed in this paper for the estimation of the
variance of the random effects
(i.e., the parameters of the matrix $V_p$).
These parameters are usually estimated by maximum likelihood techniques
[see \citet{retmen2003}, among others]
and the corresponding information matrix of these estimates is of a
fundamentally different structure compared
to the variance matrix of the least squares estimate. For the two
optimal designs
we have calculated the
$D$-efficiencies for estimating the diagonal elements of the matrix
$V_p$, which are
$ 96$\% and $79$\% in the {Uzara\tsup{\textregistered}} and
{Lanicor\tsup{\textregistered}} trial, respectively.
The~designs actually used in the trial have efficiency $98\%$ and $85\%
$, while
the corresponding efficiencies of the uniform designs are $98\%$ and
$63\%$, respectively. Thus, the proposed
optimal designs for AUC-estimation
also have reasonable efficiency for estimation
of the covariance matrix of the population distribution.

Summarizing the discussion in this example, we conclude that the
equally spaced design in the
{Uzara\tsup{\textregistered}} trial is
very close to the optimal design determined by the proposed methodology,
and it is for this reason very efficient.
However, the equally-spaced designs do not always have high efficiency.
In the {Lanicor\tsup{\textregistered}} trial, the use of naively chosen
designs yields considerably less accurate estimates.
For this reason, the application of experimental
design techniques in the context of pharmacokinetics trials is strictly
recommended.

\section*{Acknowledgments}
  The~authors
would like to thank Professor Petra Thuermann, who provided the data
sets, and
Martina Stein, who typed parts of the paper with considerable technical
expertise. We would also like to thank the referees and the Associate Editor
for constructive comments on an earlier version of this paper.

\printaddresses


\begin{thebibliography}{99}
\bibitem[\protect\citeauthoryear{Aarons}{1999}]{aarons1999}
\textsc{Aarons, L.} (1999).
Software for population pharmacokinetics and pharmacodynamics.
\textit{Clinical Pharmacokinetics} \textbf{36} 255--264.

\bibitem[\protect\citeauthoryear{Atkinson}{2008}]{atkinson2008}
\textsc{Atkinson, A.~C.} (2008).
Examples of the use of an equivalence theorem in constructing optimum
experimental designs for random-effects nonlinear regression models.
\textit{J.~Statist. Plann. Inference} \textbf{138} 2595--2606.
\MR{2439971}

\bibitem[\protect\citeauthoryear{Atkinson and Donev}{1992}]{atkdon1992}
\textsc{Atkinson, A.~C.} and \textsc{Donev, A.} (1992).
\textit{Optimum Experimental Designs.}
Clarendon Press, Oxford.

\bibitem[\protect\citeauthoryear{Atkinson et~al.}{1993}]{atchheju1993}
\textsc{Atkinson, A.~C., Chaloner, K., Herzberg, A.~M.} and \textsc{Juritz, J.} (1993).
Optimum experimental designs for properties of a compartmental model.
\textit{Biometrics} \textbf{49} 325--337.

\bibitem[\protect\citeauthoryear{Bickel and Herzberg}{1979}]{bickherz1979}
\textsc{Bickel, P.~J.} and \textsc{Herzberg, A.~M.} (1979).
Robustness of design against autocorrelation in time {I}: Asymptotic
theory, optimality for location and linear regression.
\textit{Ann. Statist.} \textbf{7} 77--95.
\MR{0515685}

\bibitem[\protect\citeauthoryear{Bickel, Herzberg and Schilling}{1981}]{bickherzsch1981}
\textsc{Bickel, P.~J., Herzberg, A.~M.} and \textsc{Schilling, M.~F.} (1981).
Robustness of design against autocorrelation in time {II}:
Optimality, theoretical and numerical results for the first-order
autoregressive process.
\textit{J.~Amer. Statist. Assoc.} \textbf{76} 870--877.
\MR{0650898}

\bibitem[\protect\citeauthoryear{Biedermann, Dette and Pepelyshev}{2004}]{biedetpep2004}
\textsc{Biedermann, S., Dette, H.} and \textsc{Pepelyshev, A.} (2004).
Maximin optimal designs for a compartmental model. In
\textit{MODA 7---Advances in Model-Oriented Design and Analysis} 41--48.
Physica-Verlag, Heidelberg.
\MR{2089324}

\bibitem[\protect\citeauthoryear{Box and Lucas}{1959}]{boxluc1959}
\textsc{Box, G. E.~P.} and \textsc{Lucas, H.~L.} (1959).
Design of experiments in non-linear situations.
\textit{Biometrika} \textbf{46} 77--90.
\MR{0102155}

\bibitem[\protect\citeauthoryear{Buelga et~al.}{2005}]{buelga2005}
\textsc{Buelga, D.~S., del {M}ar Fernandez~de {G}atta, M., Herrera, E.~V.,
Dominguez-Gil, A.} and \textsc{Garcia, M.~J.} (2005).
The~{B}ateman function revisited: A critical reevaluation of the
quantitative expressions to characterize concentrations in the one
compartment body model as a function of time with first-order invasion and
first-order elimination.
\textit{Antimicrobial Agents and Chemotherapy} \textbf{49} 103--128.

\bibitem[\protect\citeauthoryear{Chaloner and Verdinelli}{1995}]{chaver1995}
\textsc{Chaloner, K.} and \textsc{Verdinelli, I.} (1995).
Bayesian experimental design: A review.
\textit{Statist. Sci.} \textbf{10} 237--304.
\MR{1390519}

\bibitem[\protect\citeauthoryear{Chernoff}{1953}]{chernoff1953}
\textsc{Chernoff, H.} (1953).
Locally optimal designs for estimating parameters.
\textit{Ann. Math. Statist.} \textbf{24} 586--602.
\MR{0058932}

\bibitem[\protect\citeauthoryear{Colombo et~al.}{2006}]{colombo2006}
\textsc{Colombo, S., Buclin, T., Cavassini, M., Decosterd, L., Telenti, A., Biollaz,
J.} and \textsc{Csajka, C.} (2006).
Population pharmacokinetics of atazanavir in patients with human
immunodeficiency virus infection.
\textit{Antimicrobial Agents and Chemotherapy} \textbf{50} 3801--3808.

\bibitem[\protect\citeauthoryear{Dette}{1995}]{dette1995}
\textsc{Dette, H.} (1995).
Designing of experiments with respect to ``standardized'' optimality
criteria.
\textit{J.~Roy. Statist. Soc. Ser. B} \textbf{59} 97--110.
\MR{1436556}

\bibitem[\protect\citeauthoryear{Dette and {O'Brien}}{1999}]{detobr1999}
\textsc{Dette, H.} and \textsc{{O'Brien}, T.} (1999).
Optimality criteria for regression models based on predicted
variance.
\textit{Biometrika} \textbf{86} 93--106.
\MR{1688074}

\bibitem[\protect\citeauthoryear{Dette et~al.}{2008}]{debrpepi2008}
\textsc{Dette, H., Bretz, F., Pepelyshev, A.} and \textsc{Pinheiro, J.~C.} (2008).
Optimal designs for dose finding studies.
\textit{J.~Amer. Statist. Assoc.} \textbf{103} 1225--1237.
\MR{2462895}

\bibitem[\protect\citeauthoryear{Fan and Chaloner}{2003}]{fanchal2003}
\textsc{Fan, S.~K.} and \textsc{Chaloner, K.} (2003).
A geometric method for singular $c$-optimal designs.
\textit{J.~Statist. Plann. Inference} \textbf{113} 249--257.
\MR{1963044}

\bibitem[\protect\citeauthoryear{Fedorov}{1972}]{fedorov1972}
\textsc{Fedorov, V.~V.} (1972).
\textit{Theory of Optimal Experiments}.
Academic Press, New York.
\MR{0403103}

\bibitem[\protect\citeauthoryear{Fedorov and Hackl}{1997}]{fedhack1997}
\textsc{Fedorov, V.~V.} and \textsc{Hackl, P.} (1997).
\textit{Model-Oriented Design of Experiments. Lecture Notes in
Statist.} \textbf{125}.
Springer, New York.
\MR{1454123}

\bibitem[\protect\citeauthoryear{Ford, Torsney and Wu}{1992}]{fortorwu1992}
\textsc{Ford, I., Torsney, B.} and \textsc{Wu, C. F.~J.} (1992).
The~use of canonical form in the construction of locally optimum
designs for nonlinear problems.
\textit{J.~Roy. Statist. Soc. Ser. B} \textbf{54} 569--583.
\MR{1160483}

\bibitem[\protect\citeauthoryear{Gaffke and Krafft}{1982}]{gaffkra1982}
\textsc{Gaffke, N.} and \textsc{Krafft, O.} (1982).
Exact {$D$}-optimum designs for quadratic regression.
\textit{J.~Roy. Statist. Soc. Ser. B} \textbf{44} 394--397.
\MR{0693239}

\bibitem[\protect\citeauthoryear{Garrett}{1994}]{garrett1994}
\textsc{Garrett, E.~R.} (1994).
The~{B}ateman function revisited: A critical reevaluation of the
quantitative expressions to characterize concentrations in the one
compartment body model as a function of time with first-order invasion and
first-order elimination.
\textit{Journal of Pharmacokinetics and Biopharmaceutics} \textbf{22}
103--128.

\bibitem[\protect\citeauthoryear{Gibaldi and Perrier}{1982}]{gibper1982}
\textsc{Gibaldi, M.} and \textsc{Perrier, D.} (1982).
\textit{Parmacokinetics}, 2nd ed.
Dekker, New York.

\bibitem[\protect\citeauthoryear{Harville}{1976}]{harville1976}
\textsc{Harville, D.} (1976).
Extension of the {G}auss--{M}arkov theorem to include the estimation
of random effects.
\textit{Ann. Statist.} \textbf{4} 384--395.
\MR{0398007}

\bibitem[\protect\citeauthoryear{Mentr{\'{e}}, Mallet and Baccar}{1997}]{menmalbac1997}
\textsc{Mentr{\'{e}}, F., Mallet, A.} and \textsc{Baccar, D.} (1997).
Optimal design in random-effects regression models.
\textit{Biometrika} \textbf{84} 429--442.
\MR{1467058}

\bibitem[\protect\citeauthoryear{P{\'{a}}zman}{1986}]{pazman1986}
\textsc{P{\'{a}}zman, A.} (1986).
\textit{Foundations of Optimum Experimental Design}.
D. Reidel, Dordrecht.
\MR{0838958}

\bibitem[\protect\citeauthoryear{Pinheiro and Bates}{1995}]{pinbat1995}
\textsc{Pinheiro, J.~C.} and \textsc{Bates, D.~M.} (1995).
Approximations to the log-likelihood function in the nonlinear mixed
effects model.
\textit{J.~Comput. Graph. Statist.} \textbf{4} 12--35.

\bibitem[\protect\citeauthoryear{Pukelsheim}{1993}]{pukelsheim1993}
\textsc{Pukelsheim, F.} (1993).
\textit{Optimal Design of Experiments}.
Wiley, New York.
\MR{1211416}

\bibitem[\protect\citeauthoryear{Retout and Mentr{\'{e}}}{2003}]{retmen2003}
\textsc{Retout, S.} and \textsc{Mentr{\'{e}}, F.} (2003).
Further developments of the {F}isher information matrix in nonlinear
mixed-effects models with evaluation in population pharmacokinetics.
\textit{Journal of Biopharmaceutical Statistics} \textbf{13} 209--227.

\bibitem[\protect\citeauthoryear{Retout, Duffull and Mentre}{2001}]{redume2001}
\textsc{Retout, S., Duffull, S.} and \textsc{Mentre, F.} (2001).
Development and implementation of the {F}isher information matrix for
the evaluation of population pharmakokinetic designs.
\textit{Computer Methods and Programs in Biomedicine} \textbf{65} 141--151.

\bibitem[\protect\citeauthoryear{Retout, Mentr{\'{e}} and Bruno}{2002}]{retmenbru2002}
\textsc{Retout, S., Mentr{\'{e}}, F.} and \textsc{Bruno, R.} (2002).
Fisher information matrix for nonlinear mixed-effects models:
Evaluation and application for optimal design of enoxaparin population
pharmacokinetics.
\textit{Stat. Med.} \textbf{21} 2623--2639.

\bibitem[\protect\citeauthoryear{Sacks and Ylvisaker}{1966}]{sackylv1966}
\textsc{Sacks, J.} and \textsc{Ylvisaker, N.~D.} (1966).
Designs for regression problems with correlated errors.
\textit{Ann. Math. Statist.} \textbf{37} 66--89.
\MR{0192601}

\bibitem[\protect\citeauthoryear{Sacks and Ylvisaker}{1968}]{sackylv1968}
\textsc{Sacks, J.} and \textsc{Ylvisaker, N.~D.} (1968).
Designs for regression problems with correlated errors; many
parameters.
\textit{Ann. Math. Statist.} \textbf{39} 49--69.
\MR{0220424}

\bibitem[\protect\citeauthoryear{Schmelter}{2007a}]{schmelter2007a}
\textsc{Schmelter, T.} (2007a).
Considerations on group-wise identical designs for linear mixed
models.
\textit{J.~Statist. Plann. Inference} \textbf{137} 4003--4010.
\MR{2368544}

\bibitem[\protect\citeauthoryear{Schmelter}{2007b}]{schmelter2007b}
\textsc{Schmelter, T.} (2007b).
The~optimality of single-group designs for certain mixed models.
\textit{Metrika} \textbf{65} 183--193.
\MR{2288057}

\bibitem[\protect\citeauthoryear{Shargel}{1993}]{shargel1993}
\textsc{Shargel, L.} (1993).
\textit{Applied Biopharmaceutics and Pharmacokinetics}.
Chapman \& Hall, London.

\bibitem[\protect\citeauthoryear{Silvey}{1980}]{silvey1980}
\textsc{Silvey, S.~D.} (1980).
\textit{Optimal Design}.
Chapman \& Hall, London.
\MR{0606742}

\bibitem[\protect\citeauthoryear{Th{\"{u}}rmann, Neff and Fleisch}{2004}]{thuermann2004}
\textsc{Th{\"{u}}rmann, P., Neff, A.} and \textsc{Fleisch, J.} (2004).
Interference of {U}zara glycosides in assays of digitalis glycosides.
\textit{International Journal of Clinical Pharmacology and
Therapeutics} \textbf{42} 281--284.

\end{thebibliography}
\end{document}